\documentclass[aps,prc,preprint,groupedaddress]{revtex4}
\usepackage[utf8x]{inputenc}
\usepackage{color,soul,ulem}

\def\Vec#1{\mbox{\boldmath $#1$}}
\def\beq{\begin{equation}}
\def\eeq{\end{equation}}

\def\beqy{\begin{eqnarray}}
\def\eeqy{\end{eqnarray}}
\usepackage{graphicx}
\usepackage{amssymb}
\usepackage{epsfig}
\begin{document}
%%%%%%%%%%%%%%%%
%
%\preprint{}
%
\title{Spin-isospin correlated configurations in complex nuclei and neutron
  skin effect in W$^\pm$ production in high-energy proton-lead collisions}
\author{M. Alvioli}
\email[]{massimiliano.alvioli@irpi.cnr.it}
\affiliation{Consiglio Nazionale delle Ricerche, Istituto di Ricerca per la 
  Protezione Idrogeologica, via Madonna Alta 126, I-06128 Perugia, Italy}
\author{M. Strikman}
\email[]{mxs43@psu.edu}
\affiliation{104 Davey Lab, The Pennsylvania State University,
  University Park, PA 16803, USA}
\date{\today}
\begin{abstract}
  We extend our Monte Carlo algorithm for generating global configurations in nuclei to include
  different spatial distributions of protons and neutrons in heavy nuclei taking into account the
  difference of spatial correlations between two protons, two neutrons and proton-neutron pairs.
  We generate configurations for $^{48}$Ca and $^{208}$Pb neutron-rich nuclei, which
  can be used in general-purpose high-energy  A(e,e$^{\prime}$p), p-A and A-A event generators.
  As an application of lead configurations, we developed an algorithm for proton-heavy nucleus
  collisions at the LHC for final states with a hard interaction in the channels where cross section
  for p-p and p-n scattering differ. Soft interactions are taken into account in the color fluctuation
  extension of the Glauber algorithm, taking into account the inherently different transverse
  geometry of soft and hard p-N collisions.
  We use the new event generator to test an interesting observation of Ref. \cite{Paukkunen:2015bwa} that
  the ratio of W$^\pm$ production rates in p-Pb collisions should significantly deviate from the inclusive
  value for peripheral collisions due to the presence of a neutron skin. We qualitatively confirm expectation
  of Ref. \cite{Paukkunen:2015bwa} though, for a realistic centrality trigger, we find the effect to be
  a factor of two smaller than the original estimate.
\end{abstract}
%
%\keywords{}
%
\maketitle
%
%_____________________________________________________________________________________________
\section{Introduction}\label{sec:intro}

It was recently pointed out that the presence of the neutron skin in heavy nuclei leads to observable
effects in proton-ion  collisions at LHC energies due to the difference of the cross sections of a number
of hard collision processes involving quarks for pp and pn scattering \cite{Paukkunen:2015bwa}
The most practical case presented by the authors is the  asymmetry of W$^+$ and W$^-$ production cross
sections. The deviations of the asymmetry from its inclusive value are larger for peripheral collisions.
Thus, a study of this ratio should provide a sensitive test of the procedures used to determine centrality
of the proton-nucleus and nucleus-nucleus high-energy collisions.

Increased accuracy of neutron skin measurements \cite{Horowitz2014,Tarbert:2013jze} allowed
comparison of measurements with state-of-the-art nuclear structure calculations \cite{Hagen2016}.
Theoretical approaches and data analysis techniques should match such accuracy. This requires using
descriptions of nuclei capable of including fine details of nuclear structure such as nucleon-nucleon
(NN) correlations, the different extent of the neutron and proton distributions, and how NN correlations
affect the distribution of the nuclear  matter. 

One widely used approach for the description of high-energy p-A and A-A collisions is the Monte Carlo
Glauber model, whose basic ingredients are a set of Monte Carlo-generated nuclear configurations and
the Glauber multiple scattering method to calculate the impact parameter dependence of individual
inelastic interactions between the nucleons belonging to colliding nuclei (proton and nucleus).
A number of nuclear configurations can be generated beforehand, for a given nucleus, and
thus details of nuclear structure can be embedded in the configurations, and the (substantial) time
needed to calculate them with the necessary accuracy can be spent only once. This approach can
be used within many existing codes, for example HIJING \cite{Deng:2010mv}, SMASH \cite{Weil:2016zrk},
Glissando \cite{Bozek:2019wyr}, the Angantyr model \cite{Bierlich:2018xfw}, and others.

We have extended our original approach for generating nuclear configurations to include, in addition
to full spin-isospin dependent NN correlations, the neutron skin effect \textit{i.e.} the different
spatial extent of the neutron and proton distributions. This can be done in principle within our
method for any nucleus, even for different isotopes of the same nucleus, provided an accurate
experimental determination for both the proton and neutron densities is available. In this work,
we introduce fully correlated configurations for two neutron-rich nuclei, namely $^{48}$Ca and $^{208}$Pb.
Our choice of nuclei is motivated by the use of $^{208}$Pb in the heavy ion program and the LHC, and
correlation studies of e-$^{48}$Ca collisions at the TJNAF.

In this paper we perform a Monte Carlo (MC) study of  the W$^+$/W$^-$ asymmetry, utilizing the newly
generated nuclear configurations and taking into account two effects neglected in Ref. \cite{Paukkunen:2015bwa}:
fluctuations of the number of collisions at a given impact parameter, and fluctuations of centrality
determinators used in the experimental studies. Overall we find that these effects reduce the deviation
of the asymmetry from its inclusive value  by a factor of two, as compared to the results of Ref.
\cite{Paukkunen:2015bwa}.

The paper is organized as follows. Section 2 describes our results for nucleon configurations in $^{48}$Ca
and $^{208}$Pb for models with uncorrelated, central-correlated and fully-correlated configurations with
built-in neutron skin effect. Section 3 describes the algorithm for generating different hard
interactions with protons and neutrons in combination with universal soft interactions. Definitions of
centrality are presented in Section 4. Our numerical results for asymmetry are presented in Section 5,
followed by conclusions in Section 6.

%_____________________________________________________________________________________________
\section{Nuclear configurations}\label{sec:configs}

In this section we describe our results for $^{48}$Ca and $^{208}$Pb configurations calculated using
an updated version of the MC code described in \cite{Alvioli:2009ab}. The original code was modified
to account for neutron skin effect, the experimental and theoretical observation that the neutron
density extends further from the center of the nucleus than the proton density. The code also
automatically accounts for short range nucleon-nucleon (NN) correlations effects. Such effects
were explicitly investigated using correlated configurations in high-energy heavy-ion collisions
in Refs. \cite{Alvioli:2011sk,Blaizot:2014wba,Alvioli:2010yk}.

\begin{figure}[!ht]
  \centerline{\includegraphics[width=0.8\textwidth]{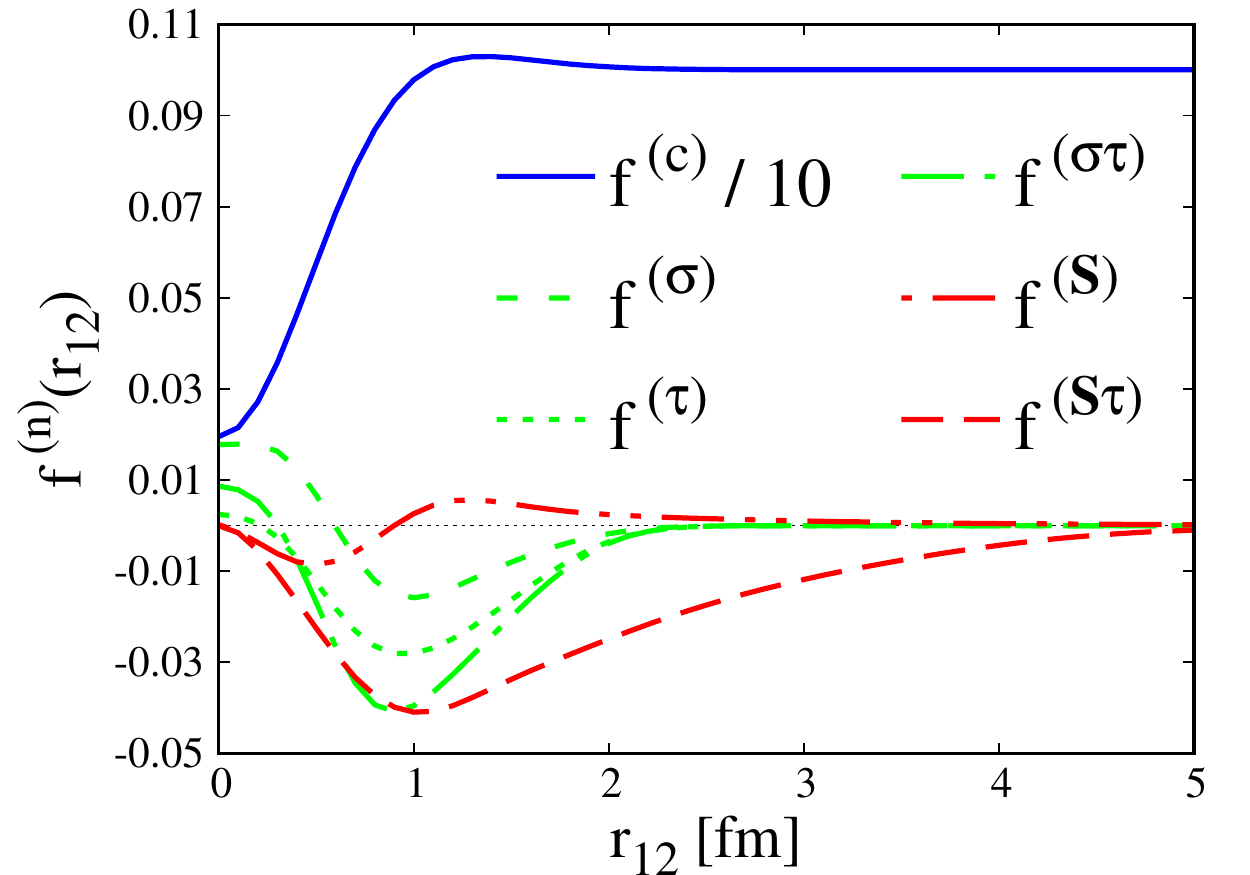}}
  \caption{The spatial dependence of the correlation functions $f^{(n)}(r_{ij}=r_{12})$
    appearing in Eq. (\ref{corrop1}). Each of the functions shown in the figure
    couples with the corresponding spin-isospin-dependent operator in Eq. (\ref{corrop1}).
    From Ref. \cite{Alvioli:2005cz}.}
  \label{fig00}
\end{figure}

The inclusion of the nucleon-nucleon correlations is based on the notion of a nuclear wave function 
$\psi$, which contains nucleonic degrees of freedom and which is used in our algorithm to modify
iteratively the positions of randomly distributed nucleons using the Metropolis method, so that
the final positions correspond to the probability density given by $|\psi|^2$. The method reproduces
single particle nucleon densities \cite{Alvioli:2009ab,Alvioli:2005cz} given by the
nucleus profile provided as an input, by construction, as well as the basic features of the two-nucleon
density \cite{Forest:1996kp,Alvioli:2007zz,Schiavilla:2006xx,Feldmeier:2011qy,Alvioli:2011aa,
  Alvioli:2013qyz,Alvioli:2016wwp,Lonardoni:2017egu,Lonardoni:2018sqo,CruzTorres2018}, calculated
accounting for NN correlations within a number of high-precision approaches. The model wave function
is taken in the following form:
\beq
\label{psi}
\psi(\Vec{x}_1,...,\Vec{x}_A)=\prod^A_{i<j}\hat{f}_{ij}\,\phi(\Vec{x}_1,...,\Vec{x}_A)\,,
\eeq
where $\phi$ is the uncorrelated wave function and $\hat{f}_{ij}$ are nucleon-nucleon
correlation operators \cite{Alvioli:2005cz}; here, $\Vec{x}_i$ denotes the position ($\Vec{r}_i$),
spin and isospin projections ($\sigma_{zi}$ and $\tau_{zi}$, respectively) of the $i$-th nucleon.
The correlation operator contains a detailed spin-isospin structure, which is the same as the
one contained in NN potentials of the Argonne family and others, which is defined as follows:
\beq
\label{corrop1}
\hat{f}_{ij}\,=\,\sum^6_{n=1}\,\hat{O}^{(n)}_{if}\,f^{(n)}(r_{ij})\,.
\eeq
Here $\hat{O}^{(n)}_{ij}$ are the standard operators \cite{Pandharipande:1979bv} used in the above
mentioned NN potentials:
\beq
\label{corrop2}
\hat{O}^{(n)}_{ij}\,=\,(1,\,\Vec{\sigma}_i\cdot\Vec{\sigma}_j,\,\Vec{S}_{ij})
\otimes(1,\,\Vec{\tau}_i\cdot\Vec{\tau}_j).
\eeq
The spatial dependence of the correlation functions $f^{(n)}$ in Eq. (\ref{corrop1}),
used in this work, is shown in Fig. \ref{fig00}.

One-body density \cite{Alvioli:2009ab,Alvioli:2005cz,Alvioli:2013qyz,Alvioli:2016wwp}
is defined as:
\beq
\label{oneb}
\rho^{(1)}(r)\,=\,\rho^{(1)}(\Vec{r}_1)|_{r_1=r}\,=\,A\,\int\prod^{A}_{i=2}
d\Vec{r}^2_i\,\left|\Psi(\Vec{r}_1,..,\Vec{r}_A)\right|^2
\eeq
and two-body density as:
\beq
\label{dummy}
\rho^{(2)}(\Vec{r}_1,\Vec{r}_2)\,=\,A(A-1)\,\int\prod^{A}_{i=3}
d\Vec{r}^2_i\,\left|\Psi(\Vec{r}_1,..,\Vec{r}_A)\right|^2\,.
\eeq
The densities in Eqs. (\ref{oneb}) and (\ref{dummy}) are spin-isospin summed quantities. If the summations
(not shown in Eqs. (\ref{oneb}) and (\ref{dummy})) over the individual isospin variables of particle ``1'',
in Eq. (\ref{oneb}), and of particles ``1'' and ``2'', in Eq. (\ref{dummy}), are not carried out, partial
quantities can be obtained. In particular, we can investigate the proton and neutron contributions to the
one-body density, and the different proton-proton, proton-neutron and neutron-neutron contributions to the
two-body density. In particular, we consider the radial two-body density:
\beq
\label{twob}
\rho^{(2)}(r_{12})\,=\,A\,\int d\Vec{R}\,
\rho\left(\Vec{r}_1=\Vec{R}+\frac{1}{2}\Vec{r},\Vec{r}_2=\Vec{R}-\frac{1}{2}\Vec{r}\right)\,.
\eeq
The quantities presented in Eqs. (\ref{oneb}) and (\ref{twob}) can be calculated  straightforwardly using  
the nuclear configurations.

We produced configurations using three different approximations, namely i) the no-correlation
approximation, ii) a repulsive, central correlation function, iii) a realistic set of spin-
and isospin-dependent correlation functions, obtained using variational calculations of medium-heavy
nuclei \cite{Alvioli:2007zz}. The approximation i) is provided as a baseline, and it can be
achieved simply by imposing that the one-body density calculated from the MC configurations
reproduces a Woods-Saxon parametrization of the nucleus profile. The approximation of ii)
was already introduced in Ref. \cite{Alvioli:2009ab}, and it can be achieved by introducing the
additional constraint that the produced configurations maximize the objective function, the square
of Eq. (\ref{psi}), where the only central correlation, $f^{(n=1)}(r_{ij}) = f^{(c)}(r_{ij}) = 1-e^{-0.9\,r^2}$
\cite{Alvioli:2009ab}, is retained in Eq. (\ref{corrop1}).
The approximation iii) was not implemented in the original version of our MC code \cite{Alvioli:2009ab}
and it was partially implemented in a previous study of initial-state anisotropies in heavy-ion collisions
from the Monte Carlo Glauber (MCG) model \cite{Alvioli:2011sk}. In this paper we present results for fully
correlated nuclear configurations, obtained by introducing NN correlations generated by including up to
the tensor, spin-isospin-dependent operator in Eq. (\ref{corrop1}). This way we effectively take into
account the three-body-induced correlations, arising from the non-commutative nature of the tensor operator
which only survives in the operator chains including three particles. A nice discussion of this effect and
a graphical representation of the tensor operator acting on three nucleons was presented in Ref. \cite{Feldmeier:2011qy}.

Inclusion of neutron skin effects in the nuclear configurations required a different parametrization
for the neutron and proton densities. Each configuration is generated producing the position of A nucleons,
distributed with a density $\rho(r)$ described by Woods-Saxon distributions with different parameters
for protons and neutrons.

For the $^{48}$Ca nucleus, we use here the parametrization of Ref. \cite{Alkhazov:1978et} for charge (proton)
and neutron densities. The parametrization of Ref. \cite{Alkhazov:1978et} has the three-parameter Fermi model
form:
\beq
\label{neutron4048}
\rho(r)\,=\,\frac{\rho_o\left(1\,+\,w\,r^2_{p,n}/c^2_{p,n}\right)}{1\,+\,e^{(r_{p,n}-c_{p,n})/z_{p,n}}}\,,
\eeq
where $w$ = -0.08, $c_p$ = 3.81 fm, $c_n$ = 4.12 fm, $z_p$ = 0.53 fm and $z_n$ = 0.51 fm, and
$\rho_{0}$ is the density at the center of the nucleus.

\begin{figure}[!ht]
  \centerline{\includegraphics[width=0.8\textwidth]{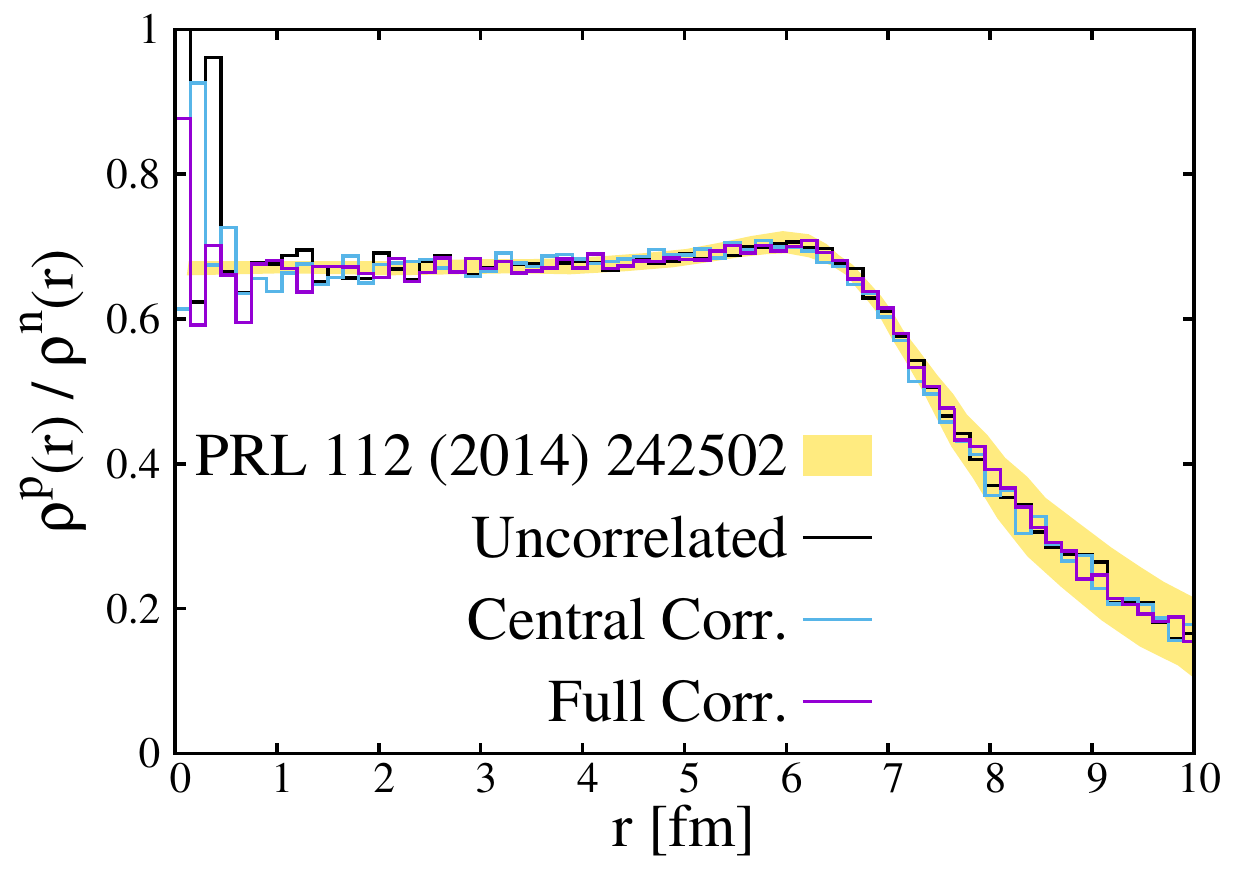}}
  \caption{The proton-to-neutron ratio of one-body densities as defined in Eq. (\ref{oneb}),
    compared with the experimental data from Ref. \cite{Tarbert:2013jze}}.
  \label{fig02}
\end{figure}

For the $^{208}$Pb nucleus, we followed the parametrization of Ref. \cite{Tarbert:2013jze}, which
has the following form (a similar approach was recently adopted in Ref. \cite{Loizides:2017ack}):
\begin{equation}
  \rho^{(p,n)}(r) = \frac{\rho_0}{1+e^{(r-R^{p,n}_0)/a_{p,n}}}\,.
  \label{eq:stdWS}
\end{equation}
The neutron radius $R_0$ and skin depth $a_n$ Woods-Saxon parameters ($R^n_0$ = 6.7 fm,
$a_n$ = 0.55 fm) were obtained in Ref. \cite{Tarbert:2013jze} using coherent pion photoproduction
data while the proton ones ($R^n_0$ = 6.68 fm, $a_n$ = 0.447 fm) are commonly taken from high-energy
elastic electron scattering measurements \cite{Warda:2010qa}.
Results for the one-body density of $^{208}$Pb are shown in Fig. \ref{fig02}. The figure shows
a comparison of the ratio of the proton one-body density, $\rho^{(p)}(r)$, to the neutron one-body
density, $\rho^{(n)}(r)$. The densities were calculated using our MC code, with uncorrelated,
central-correlated and fully-correlated configurations, and compared to the experimental
measurements of Ref. \citep{Tarbert:2013jze}. All of the calculated densities compare well
with the measured ratio, as they should, since the inclusion of NN correlations does not
affect the nucleus profile.

Figure \ref{fig03} shows the radial two-body densities (\textit{cf.} Eq. (\ref{twob})) for
both the considered nuclei, which we can also consider as the probability of finding a
given NN pair in the nucleus at relative distance $r_{12}$.
The different contributions from proton-proton, proton-neutron and neutron-neutron pairs
in Fig. \ref{fig03} are shown separately. The figure shows two-body distributions obtained
with the generated configurations, highlighting the striking differences between correlated
and uncorrelated configurations, including skin effect, for all the three approximations
described above. In particular, the inclusion of NN correlations results in vanishing two-body
densities at zero pair separation. Moreover, the fully-correlated density overshoots the
central-correlated one at NN separations between 1.0 and 2.0 fm. This feature is entirely
due to pn pairs, as it is evident from Figure \ref{fig03}(b).

Configurations including full two-body and three-body induced correlations, and including also
nuclear deformations where applicable, were produced for other nuclei:  $^{12}$C, $^{40}$Ca, $^{48}$Ca,
$^{63}$Cu, $^{197}$Au, $^{238}$U, which will be presented elsewhere. 
All configurations will be posted on our project webpage \footnote{URL: http://sites.psu.edu/color/}.
Configurations for $^{208}$Pb were also used in Ref. \cite{Alvioli:2019kcy} for a different purpose,
namely the study of double partonic interactions.

%_____________________________________________________________________________________________
\section{Hard trigger geometry}\label{sec:geometry}

The basic quantity calculated in the MCG approach, using the nuclear configurations
described in Section \ref{sec:configs}, is the probability of the projectile proton to experience
$\nu$ inelastic (soft) interactions with the nucleons of  the target nucleus. In particular,
for the purpose of this work, we are interested in calculating the separate contributions from protons
and neutrons in the target, which we denote as $P^{soft,(p;n)}(b,\nu)$. 

We can calculate the most general form of the probability of interaction with $\nu$ nucleons, with $N_p$
protons and with $N_n$ neutrons, as a function of both $\nu$ and of the impact parameter, and subsequently
we can single out only the $b$ dependence, as follows:
\beq
\label{prob_bi}
P^{soft,(p;n)}(b)\,=\,2\pi\,b\,\sum_\nu\,P^{soft,(p;n)}_\nu(b)\,,
\eeq
or the only $\nu$ dependence, integrating over ${\Vec{b}}$, as follows:
\beq
\label{pintb}
P^{soft,(p;n)}(\nu)\,=\,\int d\Vec{b}\,P^{soft,(p;n)}(b,\nu)\,,
\eeq
where \textit{soft} indicates that  inelastic interactions were restricted to soft ones. 

In a previous work \cite{Alvioli:2014sba}, we introduced a method to further require that an event
contained a hard interaction. Correspondingly, we calculated the probability $P^{hard}_{ev}(b,\nu)$
of having an event in which $\nu$ inelastic interactions occurred, one of which was a hard interaction,
for the scattering of a proton on a nucleus at the  impact parameter $b$.

Since the location of the hard interaction on the transverse plane is unknown, we can calculate the
cross section differential in impact parameter by taking the convolution of the generalized parton
distributions $F_g$ of the projectile and target nucleons, and then integrate over all the possible
transverse positions for each hard interaction and for each simulated p-Pb event. In each event, we
select one particular nucleon as the one experiencing the hard interaction, based on the probability
of hard interaction, which for each nucleon $j$ is obtained as:
\beq
p_j\,=\,\frac{F_g(\Vec{b}\,+\,\Vec{\rho}\,-\,\Vec{b}_j)}{\sum_k F_g(\Vec{b}\,+\,\Vec{\rho}\,-\,\Vec{b}_k)}\,,
\eeq
where the $\Vec{b}$ is the incoming proton's impact parameter, $\Vec{\rho}$ is its transverse distance
from the hard interaction point and $\Vec{b}_j$ is the $j$-th target nucleon transverse position. Figure
\ref{fig04} is an illustration of the transverse geometry.

Once one target nucleon is selected as the hard-interacting one, we calculate the number of soft-interacting
nucleons among the remaining $A-1$ nucleons in the target, and obtain the probability of events with
a hard trigger as follows:
\beq
P^{hard,(p;n)}_{ev}(\nu)\,=\,\frac{1}{A}\int d\Vec{b} d\Vec{\rho}\,\prod^A_{j=1} d\Vec{\rho}_j\,F_g(\rho)
\,\sum^A_{i=1}F_g(\rho_i)\,p(\nu;event)\,,
\eeq
where $p(\nu;event)$ is the probability that in, a specific event, $\nu$ inelastic collisions occurred,
including the hard one. We keep the dependence on the particular event here, because that is the
stage at which we integrate the position of the hard interaction over the whole transverse plane, in
each simulated event.

Figure \ref{fig05} presents the quantities $P^{soft;hard}(b,\nu)$ calculated using the method outlined
above, both in the Glauber approximation. A second method includes the effects of fluctuations of NN cross
section were first introduced by \cite{Baym:1994uh,Baym:1995cz,Blaettel:1993ah}, and implemented in the
MCG model in \cite{Alvioli:2013vk}. The implementation is straightforward as it requires
simply introducing the probability distribution over the strength of the p-N interaction \cite{Alvioli:2013vk}.
Various aspects of fluctuations in p-A and A-A collisions using our configurations at LHC and RHIC energies
were investigated in \cite{Alvioli:2014sba,Alvioli:2013vk,Rybczynski:2013yba}.

The various quantities in Fig. \ref{fig05} depend only on $\nu$, as they were integrated over the impact
parameter as in Eq. (\ref{pintb}), and both averaged over a significant number of events. The figure illustrates
the effect of CF on the probability distributions as a function of the number of collisions $\nu$. Both the
distributions in Fig. \ref{fig05}(a) were obtained with the standard MCG model, with fixed p-N cross section,
while the distributions in Fig. \ref{fig05}(b) were obtained including an event-by-event fluctuating p-N cross
section $\sigma^{pN}_{in}$, \textit{i.e.} with account of CF effects. It is evident that the the distributions
including CF extend to much larger values of $\nu$, as a consequence of the smearing of centrality due to the
event-by-event fluctuation of the p-N cross section \cite{Alvioli:2013vk}.

We calculate the probability that the projectile experiences one hard interaction in an event containing
a total of $\nu$ interactions. By construction, $\nu-1$ of them are soft interactions. We can distinguish
these quantities for proton and neutrons, that is, distinguish when the hard interaction occurred with a
proton or with a neutron in the target.
Figure \ref{fig06} shows the proton-to-neutron ratio of $P^{hard}(\nu) = <\,\int d\Vec{b}\,P^{hard}_{ev}(b,\nu)\,>$
distributions. In the figure,
we show quantities calculated with: i) the Glauber approximation and un-correlated configurations, ii)
Glauber and fully correlated configuration, and iii) Glauber and CF, with un-correlated configurations.
We can see that CF effects are about 10\% in the most peripheral events, while correlations effects are
rather small and go in the opposite direction. In the following, we will investigate these features in
individual centrality bins, first for the proton-to-neutron ratio, and eventually for the W$^+$/W$^-$
cross sections ratio.

The probabilities defined in Eq. (\ref{pintb}) can be integrated in the intervals of centrality
calculated as in Eq. (\ref{defbi}), for events with a hard trigger, \textit{i.e.}:
\beq
P^{hard}_{ev,i}(\nu)\,=\,\int^{b_{i+1}}_{b_{i}} d\Vec{b}\,P^{hard}_{ev}(b,\nu)\,,
\eeq
and then calculate the average number of collisions, in each centrality bin, as follows:
\beq
\label{defavei}
<\nu^{p,n}>_i\,=\,\frac{\sum_\nu\,\nu\,P^{hard(p)}_{ev,i}(\nu)}{\sum_\nu\,P^{hard(n)}_{ev,i}(\nu)}\,.
\eeq
Note that in Eq. (\ref{defavei}) we have distinguished the cases when the hard interaction
occurred with a proton or with a neutron, so that we can calculate the ratio
\beq
\label{hardpnratio}
<\nu^{p}>_i\,\,/\,\,<\nu^{n}>_i\,.
\eeq

To estimate the ratio of the W$^+$ to W$^-$ production we need to take into account that the corresponding
cross sections depend on the quark content of the nucleons. Namely, W$^+$ production on neutrons occurs
with a probability $a$, relative to W$^+$ production on protons, and vice-versa for W$^-$ production.
We introduced this dependence in our MCG code calculating new probabilities which incorporate different
weights for W production on protons and neutrons, \textit{i.e.} with different values of $a$ in the
definition of the relative probability.

%_____________________________________________________________________________________________
\section{Definition of centrality}\label{sec:centr}

As a first approximation, we define centrality bins with respect to impact
parameter $b$ as follows. Based on the definition of the total inelastic cross
section:
\beq
\sigma^A_{in}\,=\,\int d\Vec{b}\,\sum^A_{n=1}\,\sigma_n(b)\,,
\eeq
the k-$th$ term in the above equation being:
\beq
\label{sigmaenb}
\sigma_n(b)\,=\,2\pi\,b\,{{n}\choose{A}}\,\left(\sigma^{pN}_{in}\,T(b)\right)^n\,\left(1\,-\,\sigma^{pN}_{in}\,T(b)\right)^{A-n}\,,
\eeq
with $\sigma^{pN}_{in}=\sigma^{pN}_{tot}-\frac{{\sigma^{pN}_{tot}}^2}{4\pi B^2_o}$, we define bins in $b$, $[b_i,b_{i+1}]$, such as:
\beq
\label{defbi}
f_i\,=\,\frac{1}{\sigma^A_{in}}\,\int^{b_{i+1}}_{b_i}\,d\Vec{b}\,\sigma_n(b)
\eeq
where $f_i\,=\,\{0,0.1,0.2,0.3,0.4,0.5,0.6,0.7,0.8,0.9,1\}$ as required to compare with the results of
Ref. \cite{Paukkunen:2015bwa}.

The definition of centrality was refined following the method used in experimental analyses, using the ATLAS
experiment studies of centrality as follows. The correlation between hadron production at central rapidities
and at $−4.9 < \eta < −3.2$ in the nucleus outgoing direction in p-A collisions at $\sqrt{s}$ = 5 TeV can be
interpreted in the framework of CF \cite{Aad:2015zza} phenomena.
Due to the approximate Feynman scaling near the nuclear fragmentation region, energy conservation effects are
not expected to affect the total transverse energy, $\Sigma\,E_T$, or to be strongly correlated with the activity
in the rapidity-separated central and forward rapidities regions. This expectation is validated by a measurement
of $\Sigma\,E_T$ as a function of hard scattering kinematics in p-p collisions \cite{Aad:2015ziq}. Distributions
of $\Sigma\,E_T$ were constructed as a function of the number of participating nucleons, $\nu$ + 1. Simple Glauber
estimates of $\nu$ resulted in $\Sigma\,E_T$ distributions narrower than those observed in the data. Using the CF
approach, instead, leads to a broader $\nu$ distribution due to the $\sigma^{pN}_{in} > \langle\,\sigma^{pN}_{in}\,\rangle$
tail of the distribution for p-N inelastic cross section $P_p(\sigma^{pN}_{in})$ \cite{Baym:1995cz}, and produces overly
broad $\Sigma\,E_T$ distributions. Based on these observations, parametrization of $\Sigma\,E_T$ was built and used
to calculate the relative contributions from collisions with different $\nu$ values to the p-A centrality classes
(bins in $\Sigma\,E_T$) used by the ATLAS collaboration. Application of the $\Sigma\,E_T$ parametrization to our
case leads to the centrality classes shown in Fig. \ref{fig07}. The figure shows that a broad range of values for
$\nu$ contribute to each centrality class, as expected from the CF approach with a fluctuating p-N cross section. 

%_____________________________________________________________________________________________
\section{Results}\label{sec:results}

The ratio defined in Eq. (\ref{hardpnratio}) is shown in Fig. \ref{fig08}. The figure shows results
for centrality classes defined by both the total inelastic cross section method and using the $\Sigma\,E_T$
parametrization. Using the total inelastic cross section method, we find a result which is essentially
consistent with the analysis of Ref. \cite{Paukkunen:2015bwa}, in each centrality class. With this definition
of centrality, the Glauber and CF results practically coincide. 

At the same time using the experimental procedure for determining  centrality classes we find a significant
reduction of sensitivity to neutron skin effect. Account of CF effects leads to a further reduction of the
sensitivity. Qualitatively the reason is that the number of wounded nucleons at a given impact parameter
fluctuates quite significantly already in the Glauber model and even more so in the CF model.

We checked the effect of NN correlations on the quantity defined by Eq. (\ref{hardpnratio}). We have previously
done so for the proton-to-neutron ratio of inclusive $P^{hard,(p;n)}(\nu)$ probabilities, which are shown
in Fig. \ref{fig06}. Results for the same quantity, but integrated within different centrality bins, are shown
in Fig. \ref{fig09}. In this case we actually compared only the ratios obtained with the Glauber approach
(no CF effects) and with centrality determined by the $T(b)$ method, Eqs. (\ref{sigmaenb}) and (\ref{defbi}).
We repeated the calculation with un-correlated and with fully-correlated configurations. The comparison in
Fig. \ref{fig09} reveals little effect from the inclusion of NN correlations.

The final result of our work is illustrated in Fig. \ref{fig10}. Experimentally the asymmetry of W$^+$ and
W$^-$ production, described as follows:
\begin{equation}
  A= (d\sigma^+ - d\sigma^-)/(d\sigma^+ + d\sigma^-)\,,
\end{equation}
was measured at the LHC in pp scattering (for review and references see \cite{Barter:2016oan} with a maximal
value of $A\approx 0.26$. 
 
We show results for pretty large values of $a= d_{pn} \sigma^+/ d_{pn} \sigma^- \approx d_{pp} \sigma^-/ d_{pp} \sigma^-$,
namely $a$ = 0.2 and $a$ = 0.4, corresponding to production of W in the backward kinematics where a valence quark of
a nucleon annihilates with a sea antiquark of the projectile proton. In this kinematics $d_{pn} \sigma^+ = d_{pp} \sigma^-$
and $d_{pn} \sigma^- = d_{pp} \sigma^+$.
We find a reduction of the ratio of W$^+$ to W$^-$ production cross sections, when  centrality is accounted for
in an accurate way as well as color fluctuations. Typically, the deviation of the asymmetry from the inclusive
value $(Z + a N) / (a Z + N)$ is reduced by a factor of two.

Eventually, we explicitly investigated the effect of using completely un-correlated or fully-correlated nuclear
configurations; results are shown in Fig. \ref{fig11}. In both cases the inclusion of correlations provides little
to no difference. The effect is smaller or equal than that on the effective proton-to-neutron ratio, both in the
un-binned ratio, in Fig. \ref{fig06}, and in  the ratio classified in centrality bins, in Fig. \ref{fig09}.

%_____________________________________________________________________________________________
\section{Conclusions}

We investigated the possibility of assessing centrality in pA collisions by exploiting the existence of neutron
skin in the lead nucleus. The idea was originally suggested in Ref. \cite{Paukkunen:2015bwa}, by calculating the
dependence on centrality, and thus on neutron skin, of W$^{\pm}$ production in pA collisions. We have investigated
the same idea by including state-of-the-art accuracy calculation on many respects, introducing: (i) fully NN
correlated nuclear configurations with built-in neutron skin; (ii) event-by-event fluctuation of the pN cross
section (color fluctuations); (iii) accurate classification of centrality, following the experimental method
for the definition of centrality bins, instead of the purely theoretical definition based on nuclear thickness,
$T(b)$; (iv) a detailed trigger mechanism for the hard-interacting particles.

\begin{figure}[!ht]
  \centerline{\includegraphics[width=0.5\textwidth]{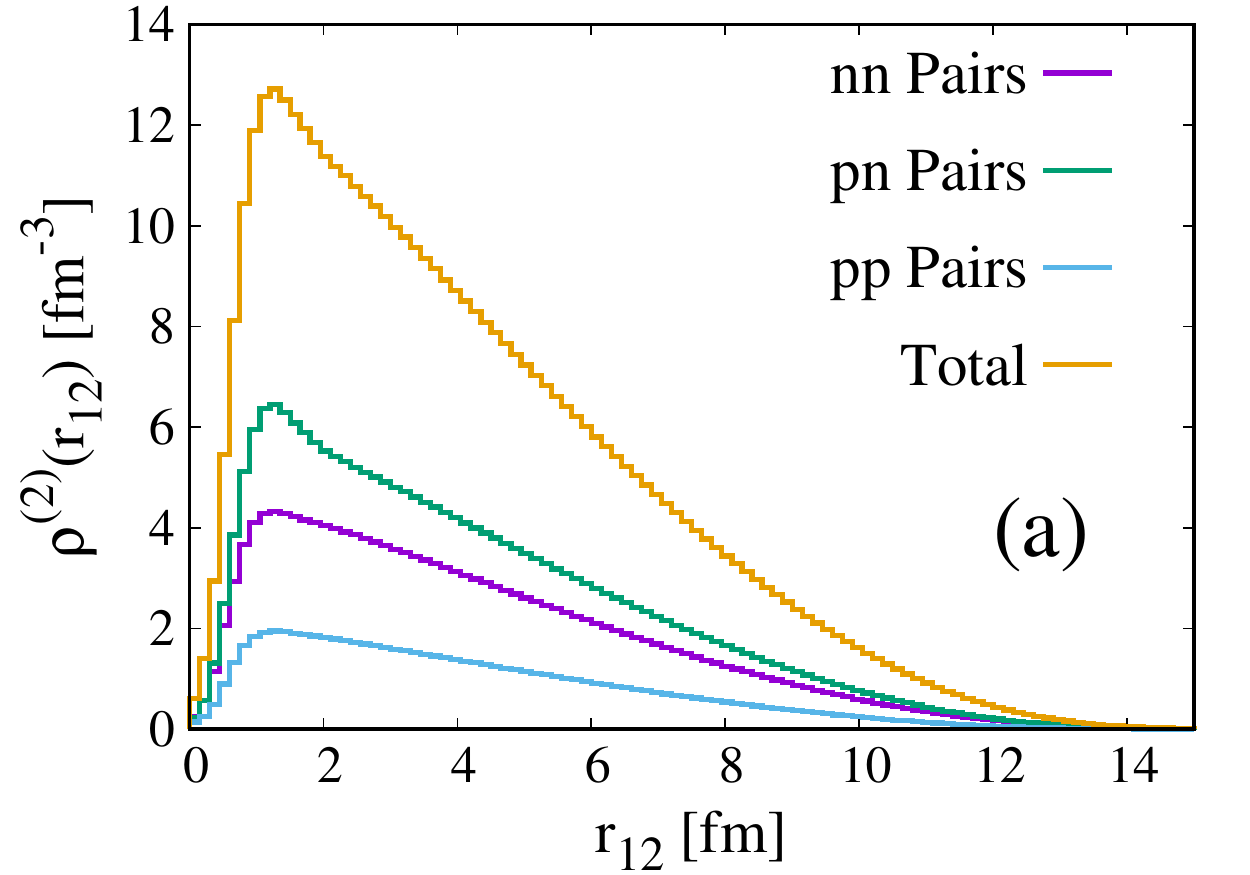}
    \includegraphics[width=0.5\textwidth]{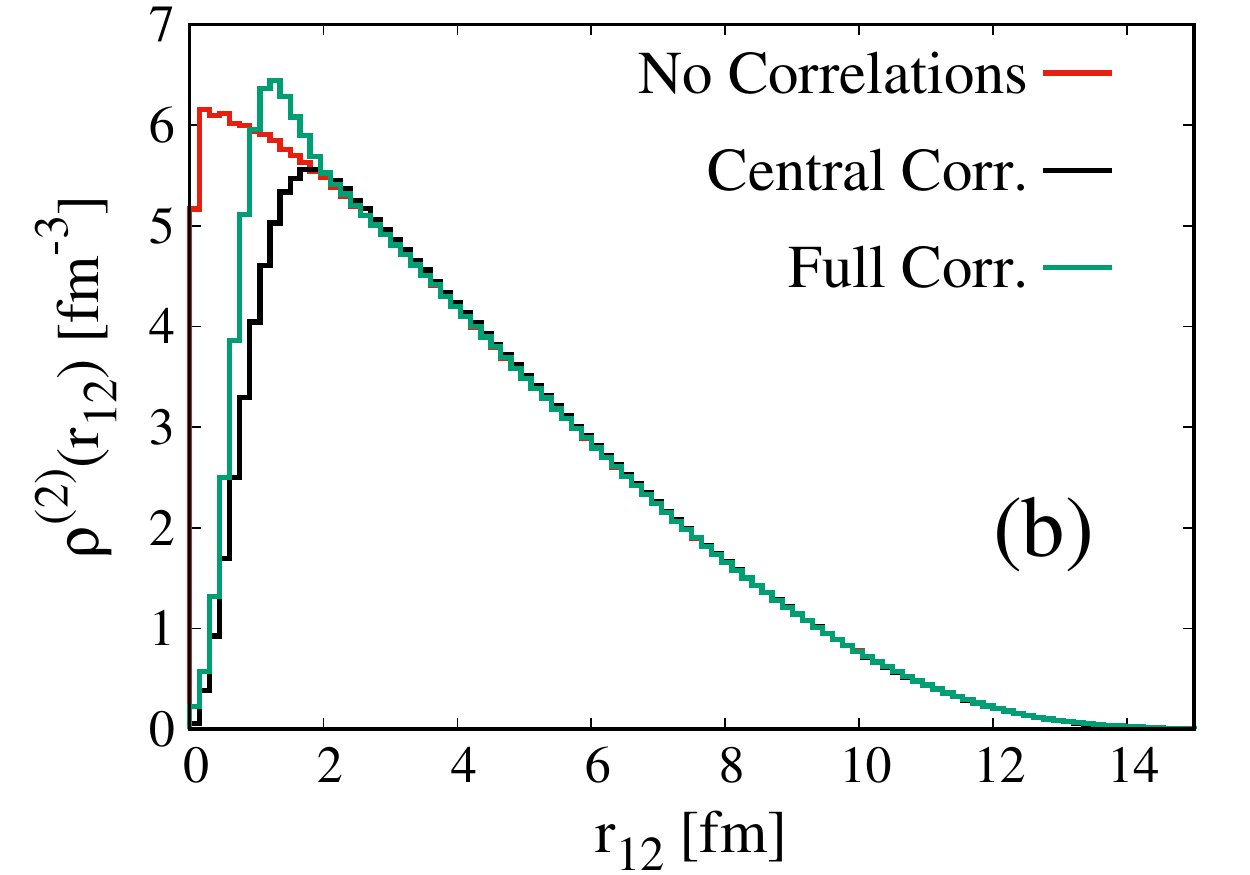}}
  \centerline{\includegraphics[width=0.5\textwidth]{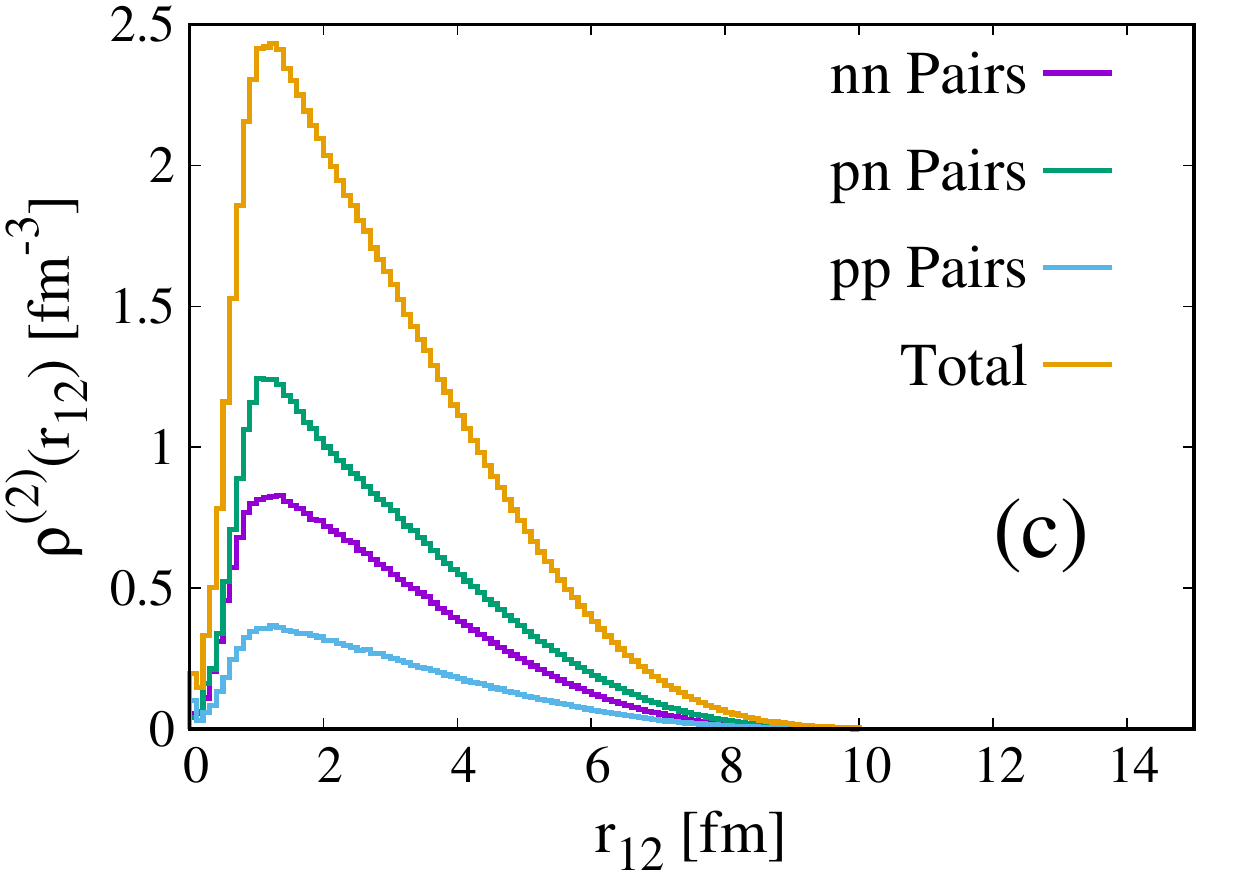}
    \includegraphics[width=0.5\textwidth]{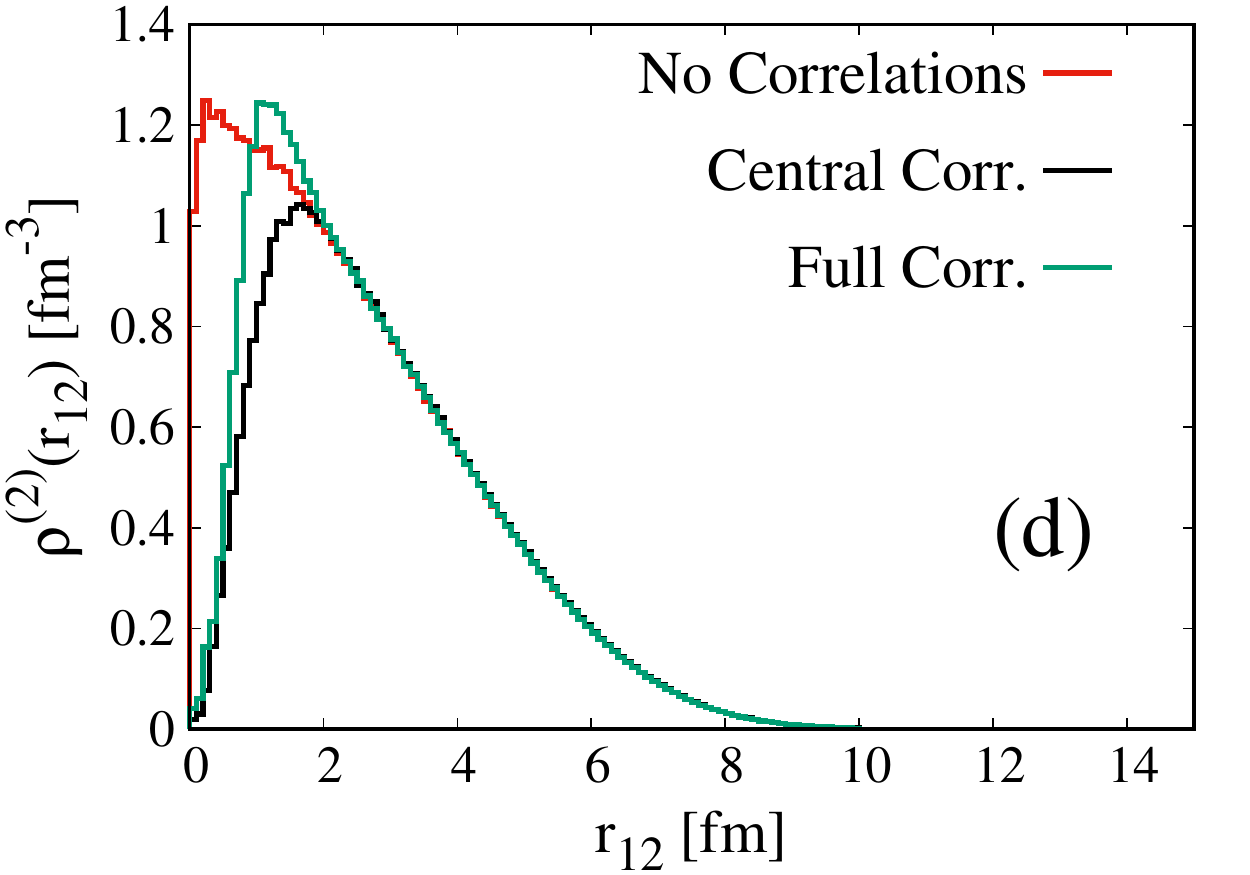}}
  \caption{The two-body density of $^{208}$Pb, in (a) and (b), and of $^{48}$Ca,
    in (c) and (d) as defined in Eq. (\ref{twob}), obtained with our configurations.
    (a) and (c): curves corresponding to different nucleon pairs, proton-proton (pp),
    proton-neutron (pn) and neutron-neutron (nn), whose sum is the total two-nucleon
    density (Total). (b) and (d): the effect of correlations in the case of proton-neutron
    pairs. All the curves are normalized according to the corresponding number of pairs
    in the nucleus.}
    \label{fig03}
\end{figure}

\begin{figure}[!ht]
\centerline{
  \includegraphics[trim={0 0 8cm 0},clip,width=0.8\textwidth]{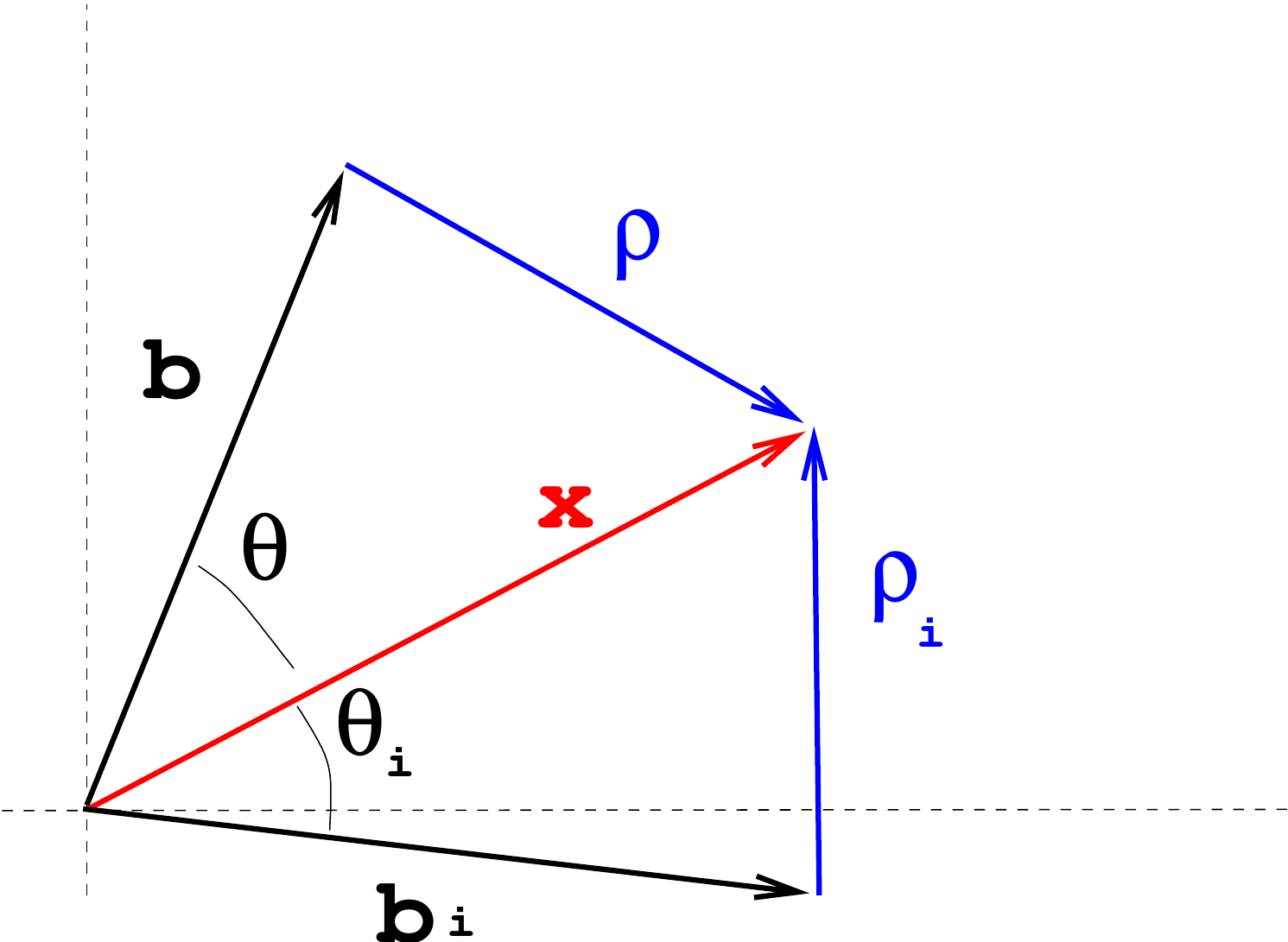}}
\caption{Sketch of the transverse geometry of a hard collision, occurring at the location
  pointed by the vector $x$. The vector $\Vec{b}$ points to the position of the incoming
  proton, and $\Vec{b}_i$ to the i$-th$ target nucleon.
  From Ref. \citep{Alvioli:2014sba}.}
  \label{fig04}
\end{figure}

\begin{figure}[!ht]
  \centerline{\includegraphics[width=0.5\textwidth]{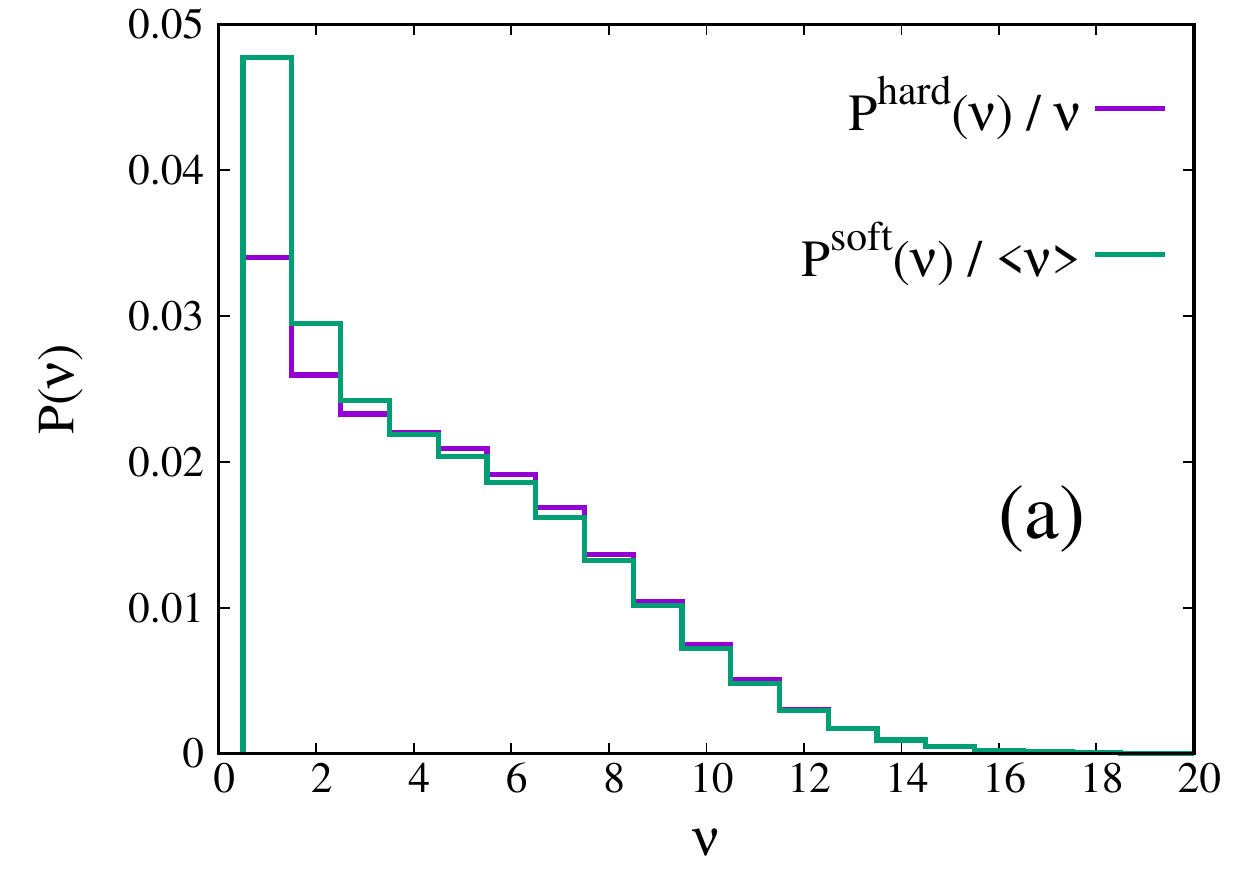}
    \includegraphics[width=0.5\textwidth]{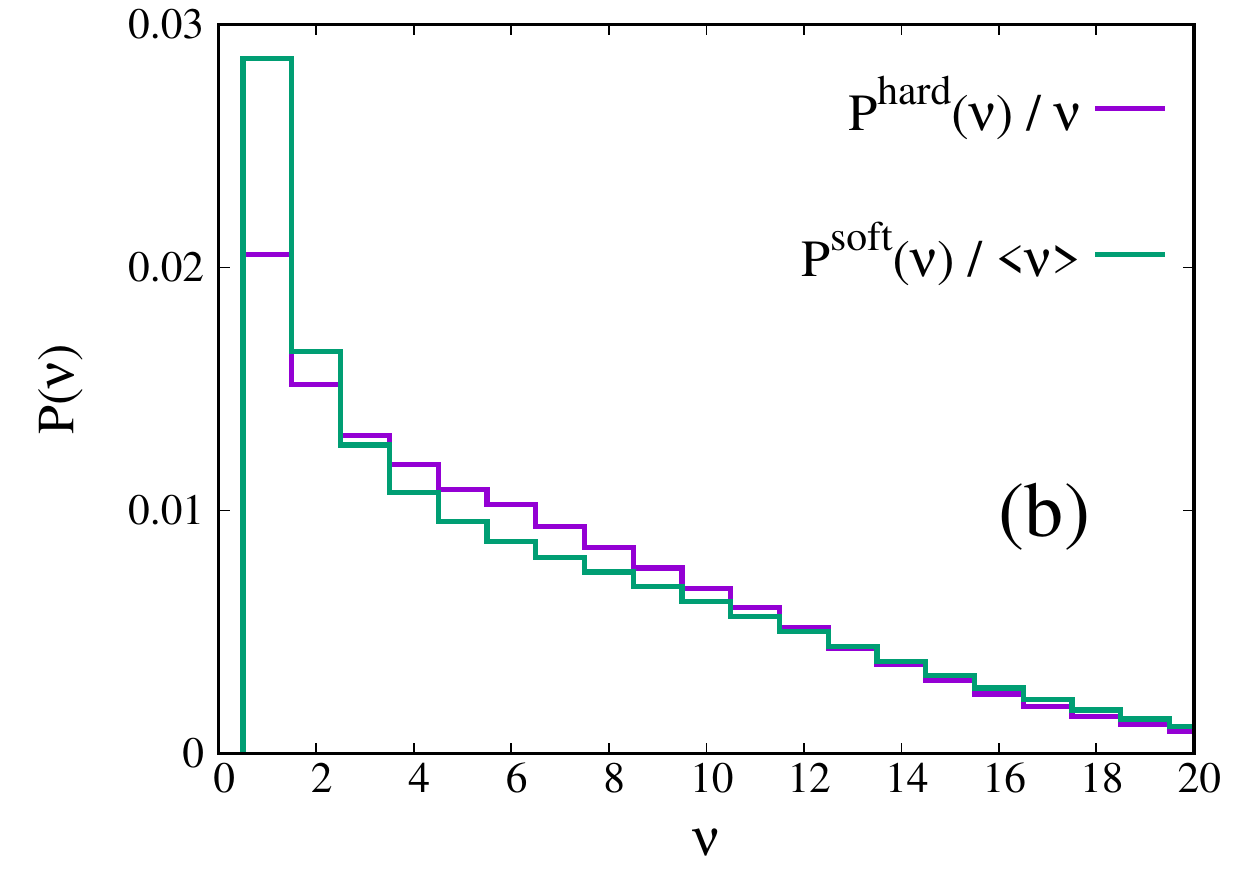}}
  \caption{Analysis of the soft and hard probabilities of interaction as a function of the
    number of collisions $\nu$. In both panels, we compare the quantity $P^{hard}(\nu)/\nu$ with
    $P^{soft}(\nu)/\langle \nu\rangle$. All calculations are independent of neutron skin effects,
    and the protons and neutrons were not distinguished, for the sake of illustration of the
    color fluctuations effect, which is absent in (a), and included in (b).}
  \label{fig05}
\end{figure}

\begin{figure}[!ht]
  \centerline{\includegraphics[width=0.8\textwidth]{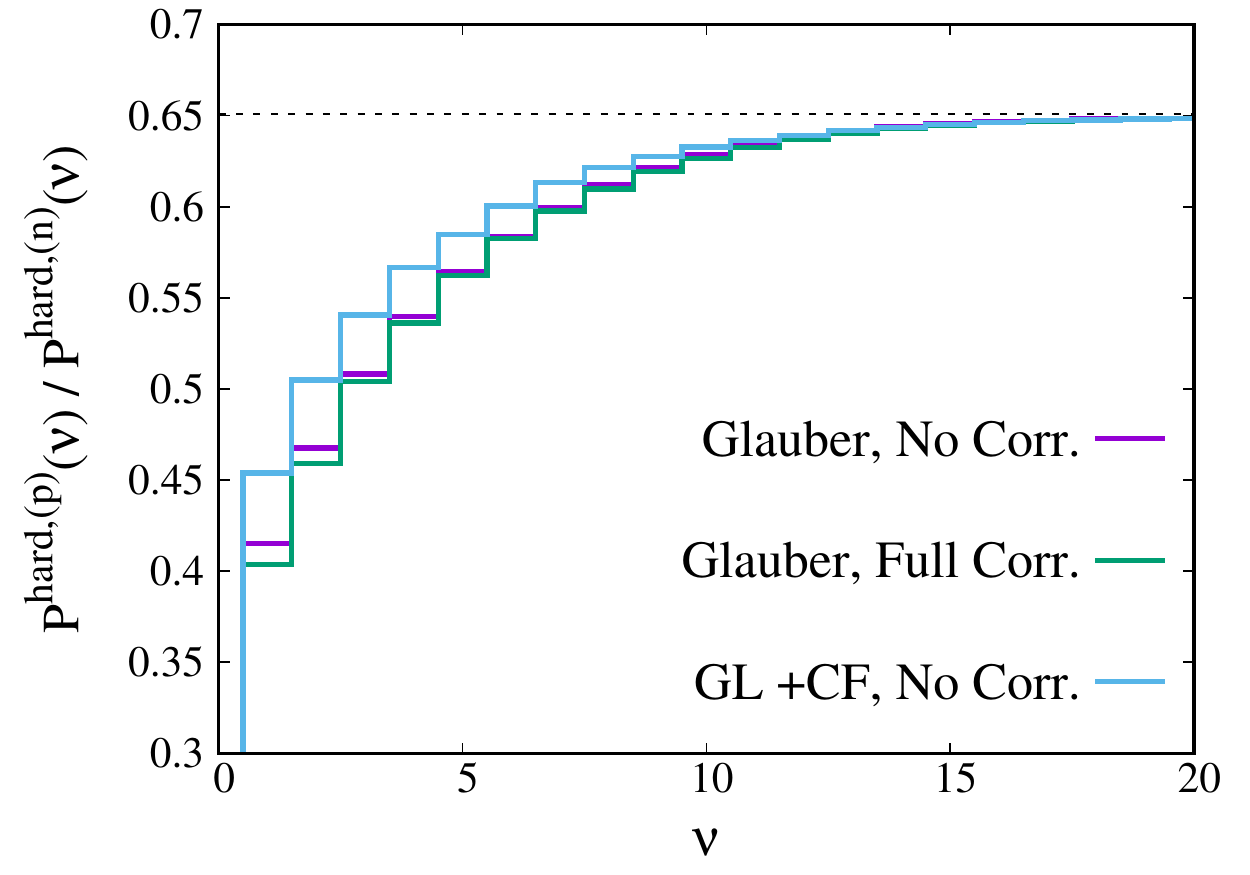}}
  \caption{The proton-to-neutron ratio of the probability
    $P^{hard}(\nu)$, calculated selecting
    only proton or neutrons in the target; we show separately the effects of NN correlations
    and of color fluctuations, for illustration purposes.
    The horizontal dashed line corresponds to $Z/N$.}
  \label{fig06}
\end{figure}

\begin{figure}[!ht]
  \centerline{
    \includegraphics[width=0.8\textwidth]{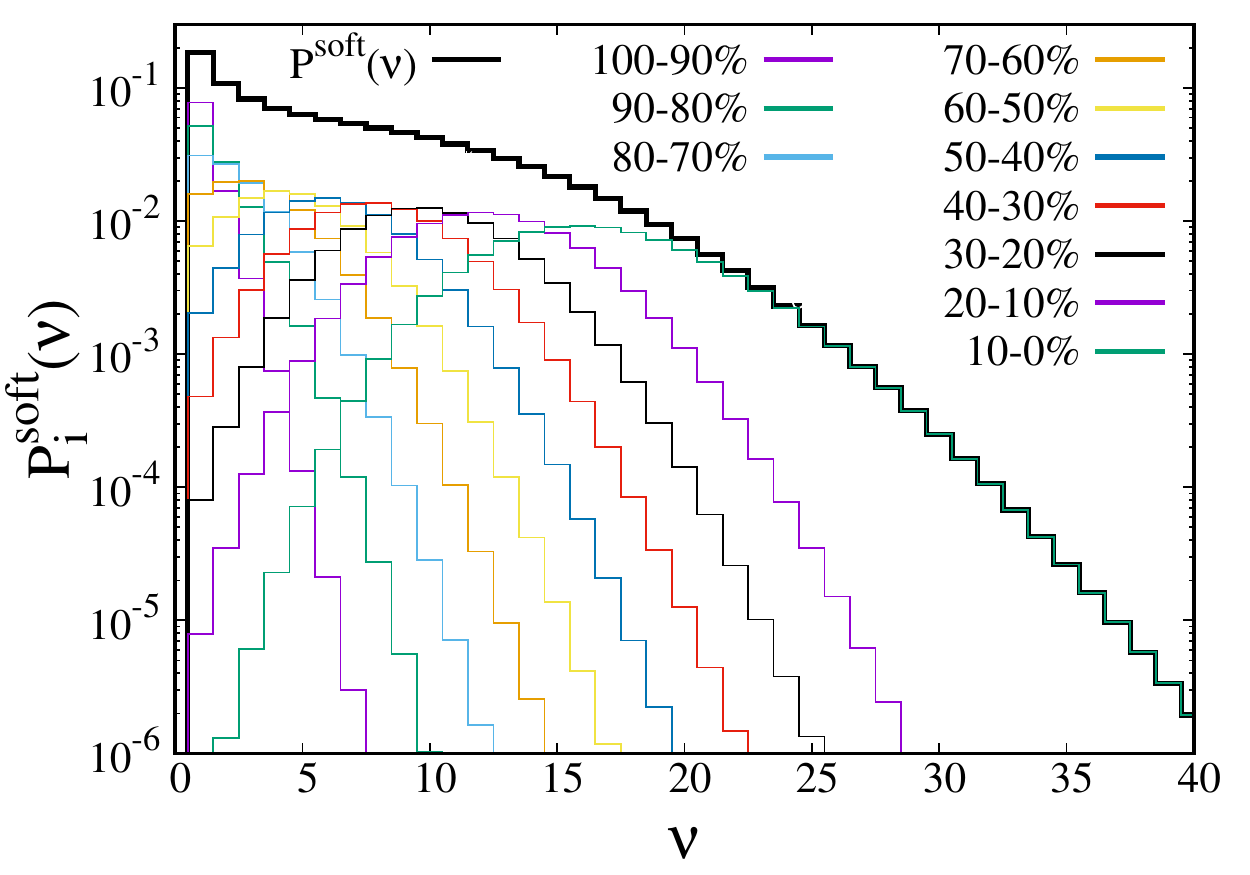}}
    \caption{The probability of interaction $P^{soft}(\nu)$, calculated within the color fluctuations
      approximation, and the different contributions to each centrality bin, according to the most
      accurate definition of centrality, based on $\Sigma\,E_T$. The thick black line shows the total
      probability; we did not distinguish between protons or neutrons, here.}
  \label{fig07}
\end{figure}

In order to ensure a realistic treatment of the nucleus wave function in modeling high-energy collisions
involving nuclei, we extended our existing event generator to produce configurations including effects
of NN correlations in different spin-isospin states and the neutron skin effect. The fully-correlated
$^{48}$Ca and $^{208}$Pb configurations show the signatures of short-range correlations \cite{Alvioli:2013qyz,
  Feldmeier:2011qy}, which are mostly found in two-body densities and can be summarized as: a) vanishing
probability of finding two nucleons at zero spatial separation, regardless of the nucleons' kind; b) the
pair probability has a maximum for 1.0 fm $\lesssim\,r_{12}\, \lesssim$ = 2.0 fm;
b) the pn probability has a more pronounced peak than pp and nn pairs, which is also found for the total
two body density (\textit{cf.} Fig. \ref{fig03}). The newly generated configurations for $^{48}$Ca and
$^{208}$Pb also include neutron skin effect, and are available for download as plain text tables, along
with configurations for other nuclei.

Configurations are are readily usable by any code which is based on Monte Carlo Glauber models
\cite{Deng:2010mv,Bozek:2019wyr} and for any kind of derived model for applications possibly
different from the one presented in this work, such as any p-A and A-A numerical model which takes
nucleon positions as an input \cite{Rybczynski:2013yba,Moreland:2014oya,Weil:2016zrk,McDonald:2016vlt},
also in combination with models for p-p studies which can be implemented within processes involving nuclei
\cite{Bierlich:2018xfw,Welsh:2016siu}.

In this work, an application of the generated configurations, in particular of the possibility of
describing nuclei using configurations with built-in NN correlations and neutron skin, is provided.
With the aim of assessing the possibility of exploiting the existence of a neutron skin in the nucleus
of lead \cite{Horowitz2014,Tarbert:2013jze}, we considered the W$^{\pm}$ production ratio in pA collisions.
We investigated separately the effects of the points (i)-(iii) above. Point (iv), consisting in the use
of an advanced hard interaction trigger for the elementary pN collisions, was used throughout the paper,
instead.

\begin{figure}[!ht]
  \centerline{\includegraphics[width=0.8\textwidth]{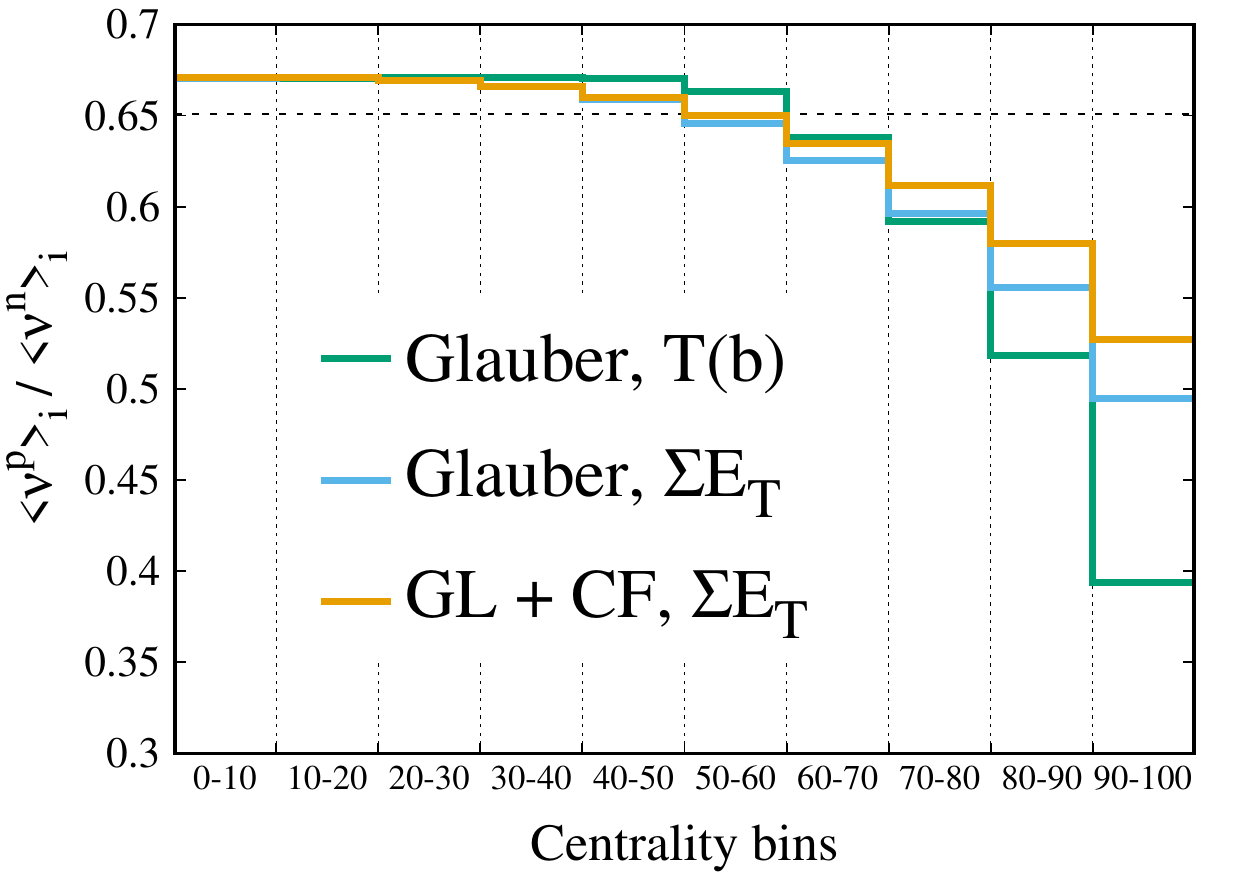}}
  \caption{The effective proton-to-neutron ratio. Green curve: the most basic approximation
    (comparable with the results of Ref. \citep{Paukkunen:2015bwa}), where we used the
    definition of centrality based on the thickness function integral, (T(b)) and Glauber
    model. Blue curve: Glauber model with definition of centrality based on experimental model
    ($\Sigma\,E_T$). Gold curve: the most refined approximation, where we used the experimental
    definition of centrality and included color fluctuations (CF) effects.
    The horizontal dashed line corresponds to $Z/N$.}
  \label{fig08}
\end{figure}

\begin{figure}[!ht]
  \centerline{
    \includegraphics[width=0.8\textwidth]{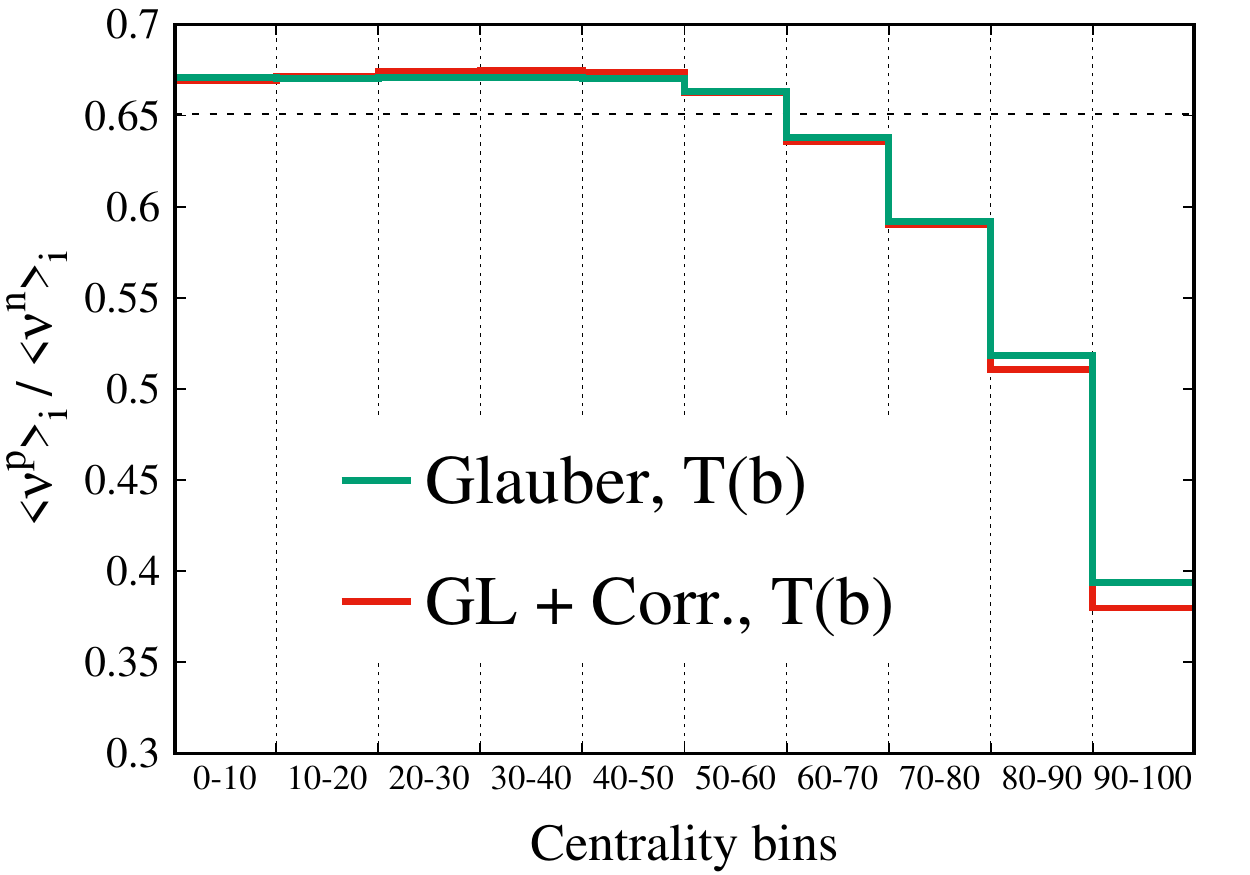}}
  \caption{Inclusion of correlated configurations on the effective
    proton-to-neutron ratio, in the simplest Glauber approximation and centrality
    defined using $T(b)$ (also shown in Fig. \ref{fig08}). The effect of NN correlations is very weak.
    The horizontal dashed line corresponds to $Z/N$.}
  \label{fig09}
\end{figure}

\begin{figure}[!ht]
  \centerline{
    \includegraphics[width=0.8\textwidth]{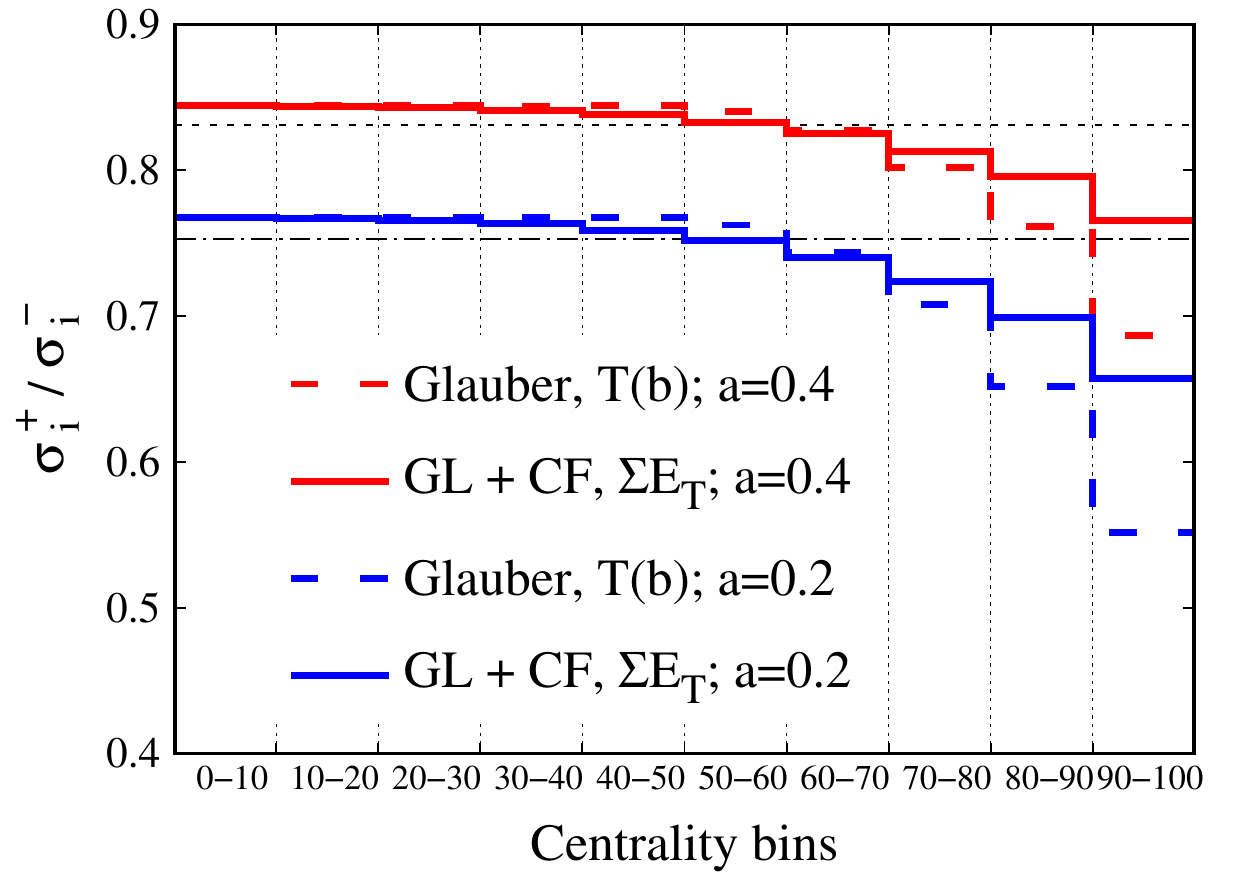}}
  \caption{The ratio of the W$^+$ production cross section to the W$^-$ one, calculated assuming
    $\sigma^{+}_i \propto P^{hard(p)}_i\,+\,a\,P^{hard(n)}_i$, and
    $\sigma^{-}_i \propto a\,P^{hard(p)}_i\,+\,P^{hard(n)}_i$, to account for the different d and u
    quark content of protons and neutrons. The dashed and dot-dashed horizontal lines correspond
    to the quantity $(Z + a N)/(a Z + N)$, for $a$ = 0.4 and $a$ = 0.2, respectively.}
  \label{fig10}
\end{figure}

\begin{figure}[!ht]
  \centerline{
    \includegraphics[width=0.8\textwidth]{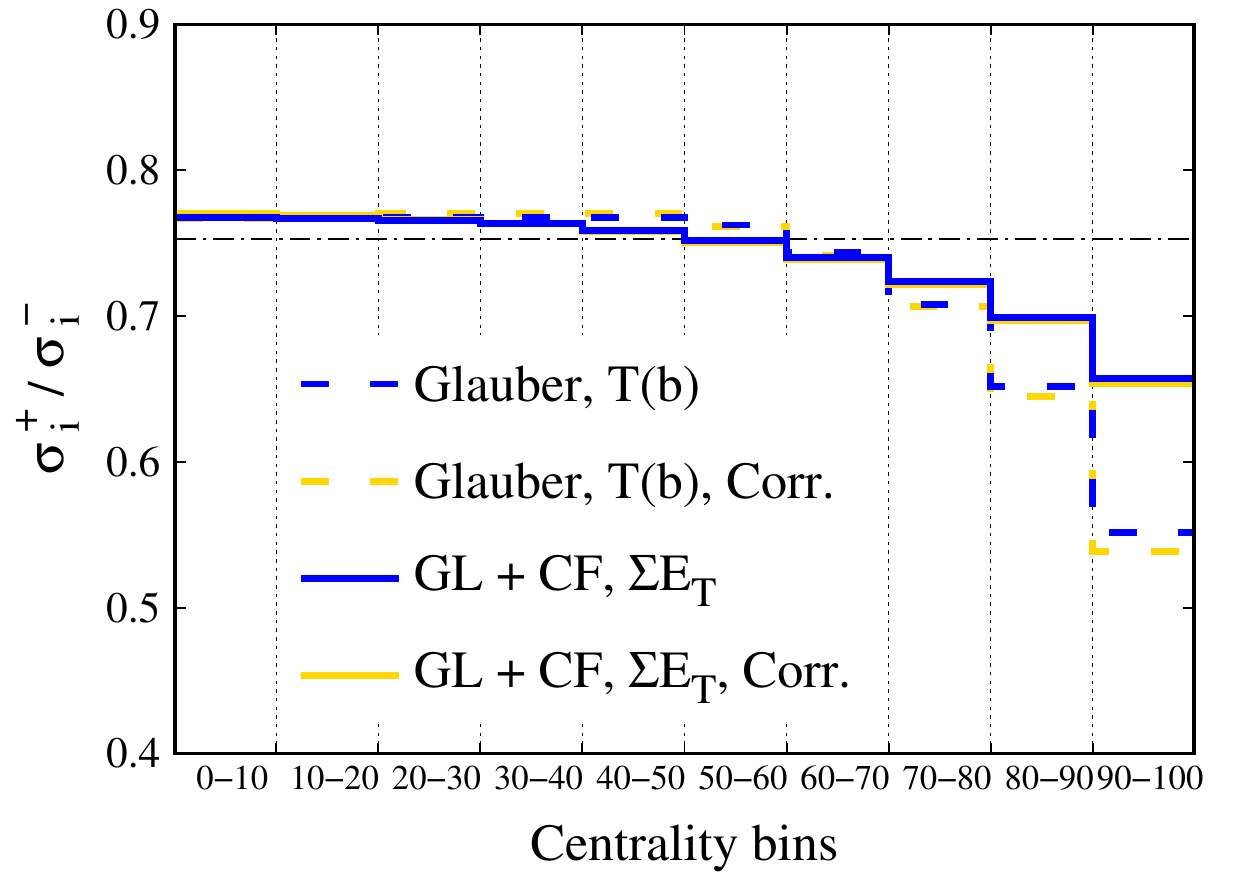}}
  \caption{Inclusion of nucleon-nucleon correlations on the ratio of the W$^+$ production cross section
    to the W$^-$ one, defined as in Fig. \ref{fig10}. We compare the results in the case $a$ = 0.2
    (blue curves, also shown in Fig. \ref{fig10}), with the corresponding calculations including
    correlations (yellow curves). Dashed lines correspond to the most basic approximation, where
    no CF nor accurate centrality definitions were accounted for; both effects are present in the
    calculation of solid lines. In both cases the inclusion of correlations provides little to no
    difference. The dashed and dot-dashed horizontal line correspond to the quantity
    $(Z + a N)/(a Z + N)$.}
  \label{fig11}
\end{figure}

We calculated results, using different approximations, for two quantities; the most basic effective
proton-to-neutron ratio (Figs. \ref{fig08}, \ref{fig09}), and the actual quantity we are interested
in, the W$^+$/W$^-$ cross sections ratio (Figs. \ref{fig10}, \ref{fig11}). In both quantities, we investigated
separately the effects of the approximations (i)-(iii) listed above. Our findings can be summarixed as follows:
\begin{itemize}
\item figure \ref{fig08} shows a comparison of the results for the proton-to-neutron reatio, as a function of centrality,
  classes, obtained according to: (a) the most basic approximation, in which centrality is accounted for using cuts in the
  integral of the nuclear thickness function $T(b)$, Eq. (\ref{defbi}), as in Ref. \cite{Paukkunen:2015bwa} (Glauber, $T(b)$);
  (b) the next approximation, in which centrality is account for using the experimental parametrization for hadronic activity
  \cite{Aad:2015zza}, as in Refs. \cite{Alvioli:2014eda,Alvioli:2017wou} (Glauber, $\Sigma E_T$); (c) the most advanced model,
  in which centrality is obtained as in (b) and color fluctuations effects are taken into account by means of event-by-event
  fluctuation of the pN inelastic cross section \cite{Alvioli:2013vk,Alvioli:2014sba,Alvioli:2014eda,Alvioli:2017wou} (GL + CF,
  $\Sigma E_T$). Results show that deviations of the ratio from its nominal value are strongly reduced in the most accurate
  estimate, with respect to the no CF, simple centrality classification method, and both the $\Sigma E_T$ classification method
  and CF effects are relevant to the result;
\item figure \ref{fig09} shows explicitly the effects of NN correlations, in the case of centrality classification
  using $T(b)$, Eq. (\ref{defbi}). The results were obtained using different nuclear configurations, either generated
  with and without inclusion of NN correlations, but including neutron skin in both cases. We can see that NN
  correlations plays little role in the effective proton-to-neutron ratio determination;
\item figure \ref{fig10} shows our estimate of the $\sigma^+/\sigma^-$ ratio, as a function of centrality classes.
  Results are presented for two values of the paramater $a$, the relative weight of production of W$^+$ from neutrons
  with respect to protons, or of W$^-$ from protons with respect to neutrons (see Section \ref{sec:results}). For both
  values of the parameters, we find that the deviation of the ratio calculated with the simplest approximation (Glauber,
  $T(b)$) is reduced by about 50\% if the most advanced approximation is used (GL + CF, $\Sigma E_T$);
\item figure \ref{fig11} shows explicitly the effect of including or not including NN correlations in the calculations
  on the $\sigma^+/\sigma^-$ ratio, as a function of centrality. The effect is shown to be negligible for both
  the simplest approximation (Glauber, $T(b)$) and for the most accurate one (GL +CF, $\Sigma E_T$).
\end{itemize}

In conclusion, we confirmed the observation of Ref. \cite{Paukkunen:2015bwa} that the ratio of the rates
of production of W$^+$ and W$^-$ in p-Pb collisions should depend on centrality of the collision due to the
presence of the neutron skin, though the inclusion of color fluctuation effects caused a reduction of the
previously predicted strength of the dependence on centrality. We also found that the expected centrality
dependence of the ratio is sensitive to the model used to determine centrality, making this process a good
testing ground for checking the centrality models especially for peripheral contributions.

Eventually, we argue that it would be possible to extend the calculation of the W$^+$/W$^-$ ratio in peripheral
Pb-Pb collisions, as in  Ref. \cite{Helenius:2016dsk},  by including the effects of fluctuations,
for an improved accuracy in modeling peripheral heavy ion collisions at collider energies.\\

{\bf Acknowledgments}\\ 

The research of M.S. was supported by the U.S. Department of Energy, Office of Science,
Office of Nuclear Physics, under Award No.  DE-FG02-93ER40771.
M.A. acknowledges a CINECA award under ISCRA initiative for making high-performance
computing resources available.\\

%%%%%%%%%%%%%%%%%%%%%%%%%%%%%%%%%%%%%%%
%%%%%%%%%%%%%%%%%%%%%%%%%%%%%%%%%%%%%%%
\newpage
%
%%---fig 01
%% end fig 01
%
\newpage
%
%%---fig 02
%% end fig 02
%
\newpage
%
%%---fig 03
%% end fig 03
%
\newpage
%
%%---fig 04
%% end fig 04
%
\newpage
%
%%---fig 05
%% end fig 05
%
\newpage
%
%%---fig 06
%% end fig 06
%
\newpage
%
%%---fig 07
%% end fig 07
%
\newpage
%
%%---fig 08
%% end fig 08
%
\newpage
%
%%---fig 09
%% end fig 09
%
\newpage
%
%%---fig 10
%% end fig 10
%
\newpage
%
%%---fig 11
%% end fig 11
%
\newpage
\bibliography{references}
%
%
%%%%%%%%%%%%%%
\end{document}